\documentclass[review]{elsarticle}
\usepackage[utf8]{inputenc}
\usepackage{siunitx}

\journal{Acta Astronautica}

\biboptions{sort&compress}

\begin{document}

\begin{frontmatter}

\title{The IPG6-B as a Research Facility to support Future Development of Electric Propulsion.}

\author[label1,label2]{Jens Schmidt\corref{mycorrespondingauthor}}
\ead{Jens\_Schmidt@baylor.edu}
\cortext[mycorrespondingauthor]{Corresponding author}
\author[label3]{Ren\'{e} Laufer}
\author[label1]{Truell Hyde}
\author[label2]{Georg Herdrich}

\address[label1]{Center for Astrophysics, Space Physics and Engineering Research (CASPER), Baylor University, 100 Research Pkwy, Waco, Texas, USA}
\address[label2]{Institute of Space Systems, University of Stuttgart, Pfaffenwaldring 29, 70569 Stuttgart, Germany}
\address[label3]{Dept. of Computer Science, Electrical and Space Engineering, Lule\r{a} University of Technology, 98128, Kiruna, Sweden}

\begin{abstract}
The inductively-heated plasma generator IPG6-B at Baylor University has been established and characterized in previous years for use as a flexible experimental research facility across multiple applications. The system uses a similar plasma generator design to its twin-facilities at the University of Stuttgart (IPG6-S) and the University of Kentucky (IPG6-UKY). The similarity between these three devices offers the advantage to reproduce results and provides comparability to achieve cross-referencing and verification. Sub- and supersonic flow conditions for Mach numbers between $Ma = 0.3 - 1.4$ have been characterized for air, argon, helium and nitrogen using a pitot probe. Overall power coupling efficiency as well as specific bulk enthalpy of the flow have been determined by calorimeter measurements to be between $\eta = 0.05 - 0.45$ and $h_s = 5- 35$ \si{\mega\joule\per\kilogram} respectively depending on gas type and pressure. Electron temperatures of $T_e = 1 - 2$ \si{\electronvolt} and densities $n_e = 10^{18} - 10^{20}$ \si{\meter^{-3}} have been measured using an electrostatic probe system. At Baylor University, laboratory experiments in the areas of astrophysics, geophysics as well as fundamental research on complex (dusty) plasmas are planned. The study of fundamental processes in low-temperature plasmas connects directly to electric propulsion systems. Most recent experiments include the study of dusty plasmas and astrophysical phenomena and the interaction of charged dust with electric and magnetic fields. In this case, dust can be used as a diagnostic for such fields and can reveal essential information of the magneto-hydrodynamics in low-temperature plasmas. Although some of these goals require further advancement of the facility, it is proposed that several phenomena relevant to electric propulsion as well as to other fields of plasma physics can be studied using the existing facility.
\end{abstract}


\end{frontmatter}

\begin{table*}\centering
	\begin{tabular}{ccl}
		\hline 
		Symbol & Unit & Description \\
		\hline 
		$Ma$ & - & Mach Number \\
		$\eta$ & -	& Overall efficiency \\ 
		$h_s$ & \si{\mega\joule\per\kilogram}	& Mass-specific enthalpy  \\ 
		$T_{e,i} $ & \si{\electronvolt}	& Electron / Ion Temperature \\ 
		$n_e$ &	\si{\per\meter^3} & Electron density\\  
		$P_{RF}$ & \si{\kilo\watt}	& Coupled RF Power \\ 
		$p_{vac} $ & \si{\pascal}	& Vacuum chamber pressure \\ 
		$p_{inj} $ & \si{\pascal}	& Injector pressure \\ 
		$p_{stat} $ & \si{\pascal}	& Static pressure \\ 
		$p_{tot} $ & \si{\pascal}	& Total pressure \\ 
		$\dot{V}_{gas}$ & SLM  & Gas Volume Flow \\ 
		$\kappa$ & - & Adiabatic Heat Coefficient \\ 
		$T_{in,out} $ & \si{\kelvin}  & In- and Outflow Water Temperature \\ 
		$\dot{V}_{cal}$ & \si{\meter^3\per\second} & Calorimeter Water Volume Flow \\ 
		$c_p $ & \si{\joule\per\kilogram\kelvin} & Specific Heat Capacity \\ 
		$I_{i,e}$ & \si{\ampere} & Electron / Ion Current \\ 
		$\varphi$ & \si{\volt} & Probe Potential \\ 	
		$L_p$ & \si{\meter} & Electrostatic Probe Length \\ 			
		$\sigma$ & \si{\siemens} & Electrical Conductivity \\
		$D_{e,i}$ & \si{\meter^2\per\second} & Electron / Ion Diffusion Coefficient\\
		$\mu_{e,i}$ & \si{\meter^2\per\volt\second} & Electron / Ion Mobility	\\
		$j_e$ & \si{\ampere\per\meter^2} & Electron Current Density \\	
		$V_p $ & \si{\volt} & Probe Voltage \\ 
		$ \epsilon $ & \si{\electronvolt} & Electron Energy \\ 
		$z, r $ & \si{\meter} & Axial and Radial Coordinates \\ 
		$P_{Cal}$ & \si{\kilo\watt}	& Calorimeter Power \\ 
		$\dot{V}_{gas}$ & \si{\kilogram\per\second}  & Gas Mass Flow \\ 
		$m_{p}$ & \si{\kilogram}  & Particle Mass \\ 
		$v$ & \si{\meter\per\second}  & Velocity\\ 
		\hline 
	\end{tabular} 
\end{table*}

\section{Introduction}

Over the past several years, the 6th generation inductively-heated plasma generator at Baylor University (IPG6-B) has been established as a research facility for application across numerous fields \cite{zhukhovitskii_electrical_2015, dropmann_new_2013-1, koch_setup_2012}. These include fundamental physics research in the field of (aerospace) engineering, astrophysics, geophysics and complex (dusty) plasma physics. Inductively-heated plasmas have been used widely for different applications such as low-power inductive laboratory discharges which are used for fundamental studies of discharge characteristics and plasma chemistry\cite{schram_fundamental_1996, godyak_electrical_1994, gudmundsson_magnetic_1997}, while high power discharges are used for materials processing or high enthalpy plasma wind tunnels for the simulation of atmospheric entry and the study of associated heat-shield materials \cite{herdrich_operational_2002, massuti-ballester_experimental_2017}. The operational regime of the IPG6-B is in between these two categories. This should allow finding new applications and operating conditions for inductively-coupled discharges, one of them being atmosphere-breathing electric propulsion (ABEP) systems as well as the study of fundamental processes in flowing plasmas with a special focus on electric propulsion,  magnetic confinement fusion and dusty plasma flows.  

\paragraph{Twin facilities}
The IPG6 is a downscaled version of the IPG3 at the Institute for Space Systems (IRS) \cite{herdrich_operational_2002} at the University of Stuttgart and is designed for lower powers in the range of $P_{RF} = 0.5 - 15$ \si{\kilo\watt}. To date, besides the IPG6-B two largely identical facilities developed in a common approach exist in two different locations: The IPG6-S at the University of Stuttgart \cite{romano_performance_2017}, used as a development platform for atmosphere-breathing electric propulsion (ABEP) systems \cite{romano_system_2018} and the IPG6-UKY at the University of Kentucky \cite{koch_operational_2016}, used to study gas-surface interactions in plasmas. These facilities are identical in generator design, e.g. discharge tube diameter, coil and achievable powers and comparable flow enthalpies. This combination provides the unique opportunity to study the behavior of similar plasma generators and facilities for different environments and diagnostics \cite{koch_optical_2017,massuti-ballester_experimental_2017,burghaus_characterization_2019}. The operation of multiple facilities also provides the advantage of allowing a common verification database and accelerated development of IPG technology due to the utilization of synergy effects and sharing of research progress between the facilities. While many other inductively coupled discharges are in operation in laboratories around the world \cite{xu_low-frequency_2001,schram_fundamental_1996,godyak_electrical_1994,kim_mass_2001,leonhardt_plasma_2001,koroglu_plasma_2017,gudmundsson_magnetic_1997,poirier_validity_2011}, which have been discussed in literature \cite{okumura_inductively_2011,lee_review_2018}, there are significant differences in plasma generator design between these facilities (e.g. frequency, coil geometry, discharge channel geometry, achievable vacuum chamber pressure) due to the fact that they were developed independently.

\begin{figure}[h]
	\centering
	\includegraphics[width=1\linewidth]{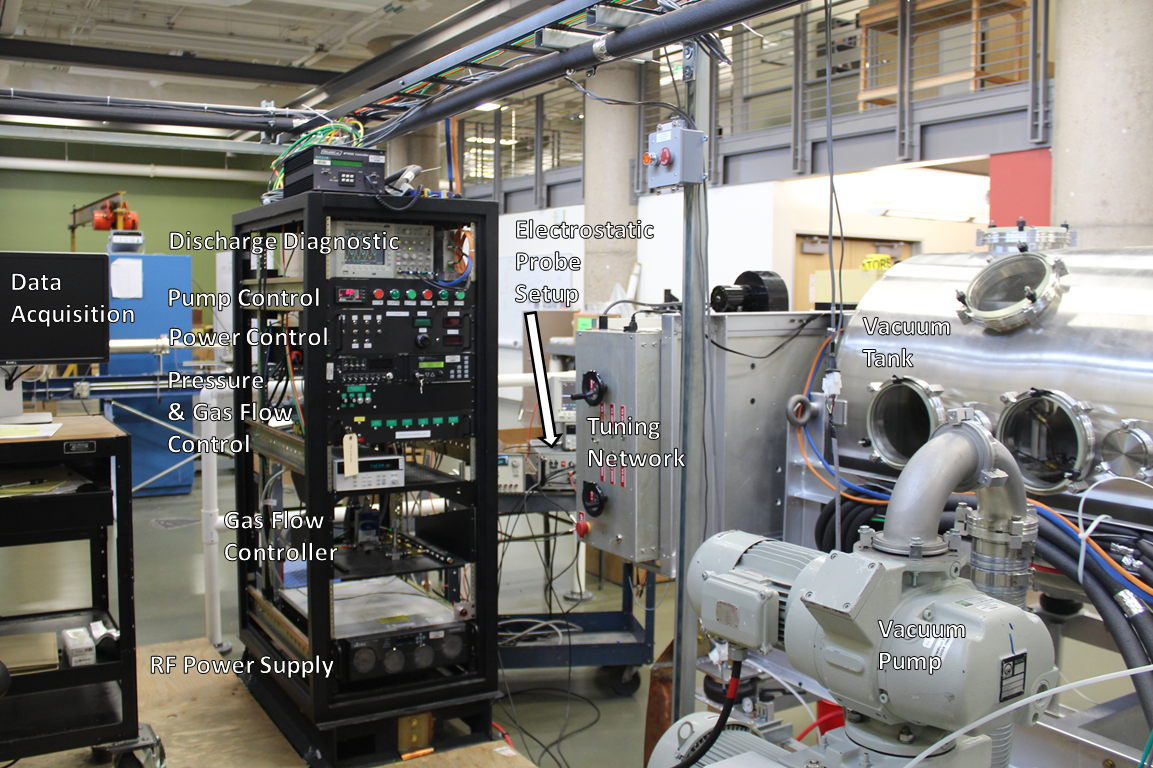}
	\caption{Image of the IPG6-B facility with description.}
	\label{fig:facilityimage}
\end{figure}

\section{The IPG6-B Facility}
The experimental facility as shown in Fig. 1 and schematically described in Fig. 2 consists of a vacuum tank and the plasma generator with tuning network and power supply. The vacuum tank has an outer diameter of 0.9 m and a length of 1.8 m leading to an overall volume of 1.2\si{\meter^3}. The vacuum tank has 11 windows, which are used for optical diagnostics. In the normative operation position, the plasma generator is mounted to one end of the tank and is connected to a Pi-type tuning network of a MKS SurePower QL 15013A Power Supply capable of supplying between 0.5 to 15kW at a frequency of 13.56 MHz. To date, the facility has been operated with air, argon, helium, oxygen and nitrogen. The plasma generator, tuning network and heat shield inside the tank are cooled using de-ionized water. 

\begin{figure}
	\centering
	\includegraphics[width=0.9\linewidth]{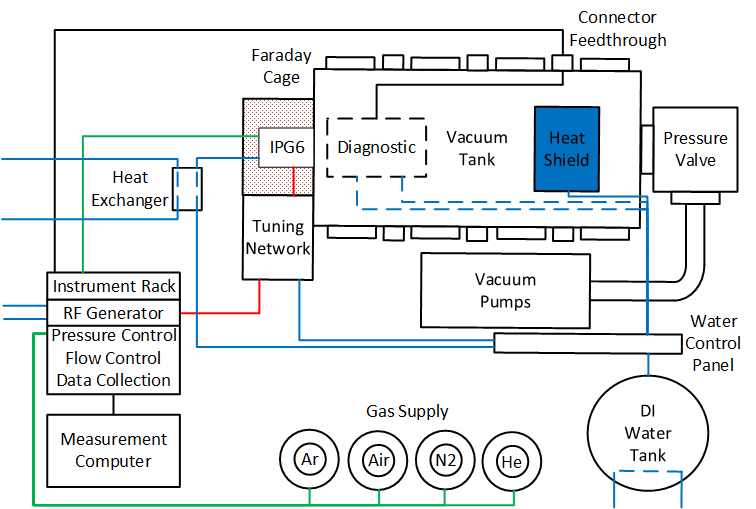}
	\caption{Scheme of the IPG6-B facility. Green lines indicate gas piping, blue lines indicate water piping, red lines indicate RF power connections and black lines indicate data transfer.}
	\label{fig:facilityscheme}
\end{figure}

\begin{figure}
	\centering
	\includegraphics[width=0.8\linewidth]{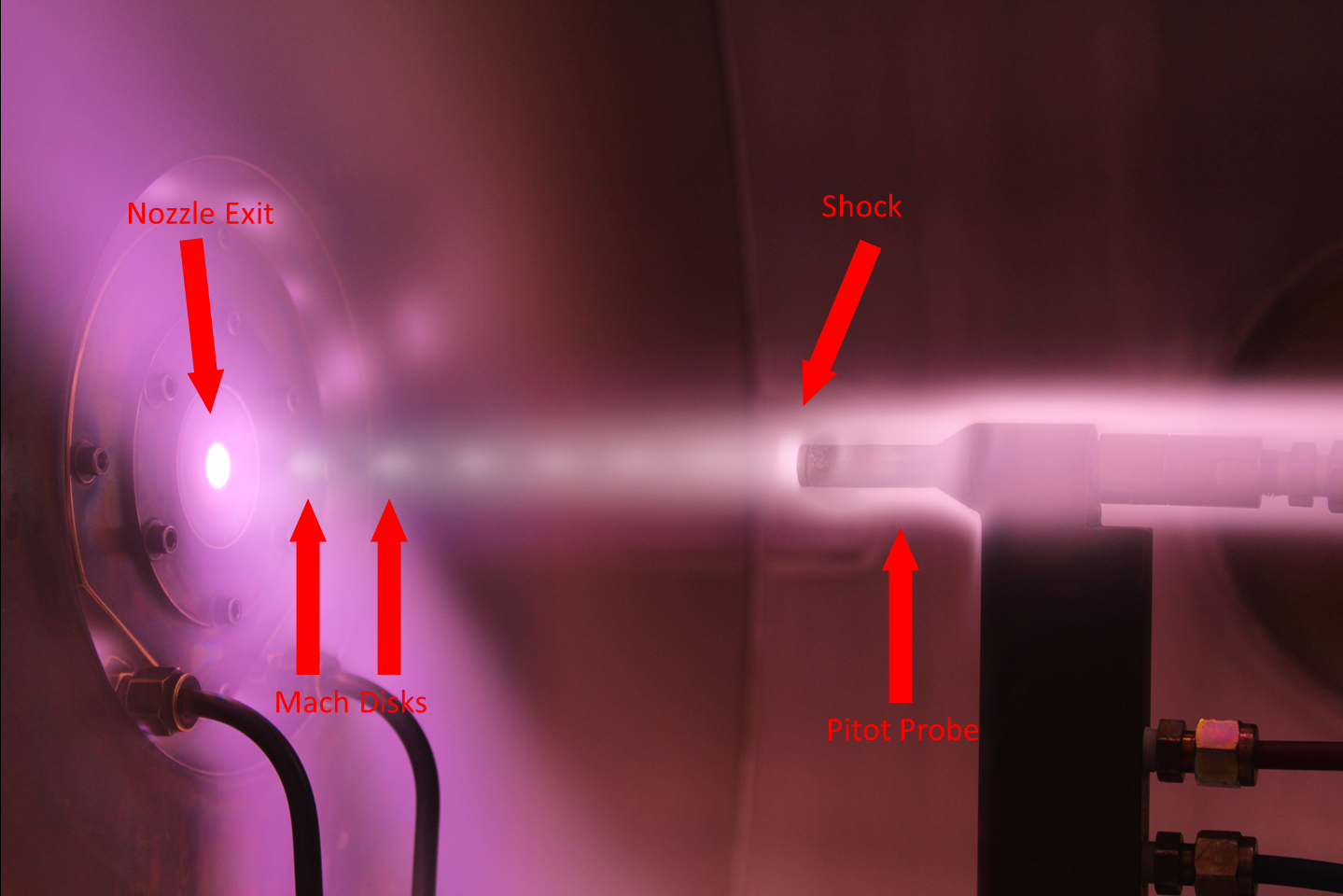}
	\caption{Picture of the facility in operation with Argon and the 10mm nozzle at $P_{RF} = 2.5$ \si{\kilo\watt} and $\dot{V}_{gas} = 6$ SLM and a Vacuum chamber Pressure of $p_{vac} = 200$ Pa during measurements using the pitot probe mounted in a water-cooled holder. The measured Mach number is $Ma = 1.4$. The Mach disks of the supersonic argon jet are clearly visible, as well as the shock in front of the pitot probe.}
	\label{fig:facilityoperation}
\end{figure}

\paragraph{The inductively-heated plasma generator IPG6-B}
The inductively-heated plasma generator is a thin walled, water-cooled generator \cite{herdrich_high-enthalpy_2008} which can be separated into 3 major parts: the gas injector, the discharge tube consisting of a water-cooled quartz-tube with a length of 180 mm and a diameter of 40 mm and a water-cooled coil with n = 5.5 windings and an exchangeable nozzle. Different convergent nozzles with nozzle throat diameters of 5, 10, 20 and 40 mm can be used. A picture of the facility in operation with Argon and a 10mm nozzle creating a supersonic plasma jet with a flow Mach number of 1.4 is shown in Fig. 3. An inductively-heated plasma generator has no electrode directly exposed to the plasma and therefore allows the operation with reactive working gases. 

\subsection{Diagnostics}

Various diagnostics have been developed for use in the IPG, including a cavity calorimeter, a pitot probe and electrostatic single and triple probes \cite{schmidt_electrostatic_2018}. Furthermore, voltage and current as well as E and H-Field are measured using an oscilloscope connected to a Rogowski coil at the conductor between the tuning network and the inductive coil of the IPG. Cooling water temperatures and volume flows are measured using thermistors and turbine flow meters, while the gas flow rate is set using a \mbox{\textit{MKS \textregistered}} mass flow controller. Forward, reflected and load RF power $P_{fwd}$, $P_{ref}$ and $P_{RF}$ can be read out at the power supply, giving information on the discharge properties. A cylindrical coordinate system with coordinate $z$ along the center axis, $r$ as the radius and $ \varphi \in [0,2\pi] $ as the angular position is defined with the exit plane of the nozzle being the origin. All diagnostics are mounted on a movable 2-axis positioning system allowing positioning along the horizontal $z$ and $r$ axes.
\paragraph{Pitot probe}
For the measurement of the total pressure of the plasma jet, a miniature pitot probe has been developed based on a similar, previously characterized probe \cite{petkow_probabilistic_2013}. A scheme of the water-cooled probe used within the IPG6-B facility is shown in Fig.  \ref{fig:facilityoperation}. It consists of the probe itself and a water-cooled probe holder, which can also be used to cool other diagnostics. The bore-hole of the probe has a diameter of \mbox{$d_{pit} = 2$ \si{\milli\meter}} while the outer diameter of the probe holder is \mbox{$d_{probe} = 6$ \si{\milli\meter}} causing only minimal disturbance to the flow. The total pressure $p_{tot}$ in front of the probe is measured. 
For subsonic flows with a pressure ratio $PR = \frac{p_{tot}}{p_{\infty}} \leq 1.8$, the Mach number $Ma$ can be directly calculated using the Rayleigh-Pitot-relation
\begin{equation}\label{eq:pitot:subsonic}
Ma = \sqrt{\frac{2}{\kappa-1} \cdot\Big[(\frac{p_{tot}}{p_{stat}})^{\frac{\kappa-1}{\kappa}}-1\Big]}
\end{equation}
under the assumption that the static pressure $p_{stat}$ is equal vacuum chamber pressure $p_{vac}$ with $\kappa$ being the adiabatic coefficient of the employed gas. The adiabatic coefficient  $\kappa$ is assumed to be constant, as it can be shown that the influence of temperature is less than 5 \% for argon\cite{artmann_uber_1964}. Furthermore, for air the difference in the resulting Mach number for $Ma \leq 2$ is less than 10\% for a variation within the range of $\kappa = 1.1 - 1.4$ \cite{herdrich_aufbau_2004}. For supersonic flow with a pressure ratio $PR \geq 1.8$ , the shock is assumed to be orthogonal to the front of the pitot probe, so the Mach number $Ma_1$ before the shock can be calculated employing
\begin{equation}\label{eq:pitot:supersonic}
\frac{p_{tot}}{p_{stat}} = \frac{\kappa-1}{2} \cdot Ma_1^2 \Big[ \frac{(\kappa+1)Ma_1^2}{4 \kappa Ma_1^2 - 2 (\kappa - 1)} \Big]^{\frac{1}{\kappa - 1}}
\end{equation}
This equation can not be solved directly for $Ma_1$, therefore the Mach-Number is iterated to match the measured pressure ratio.
\paragraph{Calorimeter}
A cavity calorimeter developed for the application in the IPG6-B \cite{dropmann_new_2013-1} and used to determine the plasma power for various working conditions by directing the plasma jet into a water cooled (copper) cavity. Under the assumption that the plasma completely recombines and de-accelerates within the calorimeter, the plasma power $P_{cal}$ can now be calculated from the temperature change $\Delta T = (T_{out} - T_{in})$ of the cooling water using \begin{equation}
P_{cal} = \dot{V}_{cal} \rho_{W} c_{p,W} \Delta T
\end{equation} where $\dot{V}_{cal}$ is the water volume flow rate, heat capacity $c_{p,W}$ and density $\rho_{W}$.  It should be noted, that for very low plasma powers, the error in the calorimeter measurement raises significantly due to the very low value of $\Delta T$.
\paragraph{Electrostatic Probe}
An electrostatic probe has been developed, consisting of 3 tungsten electrodes having a diameter of $d_{p} = 1.016$ \si{\milli\meter} contained within a water-cooled alumina rod and probe mount, movable within the 2 horizontal axes, in which the electrodes are soldered to wires connecting them to a 9-pin connector. From the probe mount a shielded cable is fed to the outside of the tank where the measurement resistor (with a resistance of $R_{meas} = 12.4 $ \si{\ohm}) for the probe current measurement is positioned. At the outside of the tank, a \mbox{\textit{Agilent 33129A}} signal generator, the \mbox{\textit{Kepco BOP 200-1M}} bipolar operational power amplifier as well as the \mbox{\textit{Tektronix TDS3014C}} digital 4-channel oscilloscope are connected. For the single/double probe setup, two electrodes are used and connected, while the third electrode is left unconnected. A potential across a range of $\pm 50$ \si{\volt} at a frequency of $f_P=250$ \si{\hertz} is applied to electrode 1 and directly measured at the oscilloscope. Electrode 2 is used as a reference electrode and directly connected to the oscilloscope. Passive filtering is provided using a ceramic capacitor with a capacitance of $C_1=15$ \si{\micro\farad}, leading to a cut-off frequency of $f_c=1$ \si{\kilo\hertz}. 
It has been found that for some working points of the IPG, the charged species are in a hydrodynamic regime which is collisional, as determined by evaluation of the Knudsen Number $Kn = \frac{\lambda_{e,i}}{d_p}$  \cite{cherrington_use_1982,demidov_electric_2002}, the ratio between the mean free path $\lambda_{e,i}$ of the charged species and the characteristic length $d_p$, in our case the probe diameter. Therefore collisional probe theory had to be applied, taking into account the electron and ion diffusion through the sheath of the probe \cite{su_continuum_1966,chen_studies_1971}. 
According to the collisional probe theory \cite{su_continuum_1966}, the relationship between probe current $I$ and probe potential $\varphi$ for a probe of length $L_p$ and radius $r_p$ is given by
\begin{equation}\label{eq:ep:1}
\frac{d I}{d \varphi} = \frac{3 \pi L_p \sigma}{ln(\frac{\pi L_p}{4 r_p})ln(\frac{r_p}{h})}
\end{equation} with the abbreviation $h = \sqrt{\frac{k_b T_e}{4 \pi e^2 n_e}}$ and electric conductivity $\sigma = n_e e^2 [\mu_e + \mu_i]$ where $\mu_{e,i} = \frac{e D_{e,i]}}{k T_{e,i}}$ are the electron and ion mobility. The electron and ion saturation currents $(I_{e,i})_{sat}$ are then given by \begin{equation}\label{eq:ep:2}
(I_{e,i})_{sat} = n_{e} \Big[ 2 \pi L_p k_b (T_e + T_i) \frac{\mu_{e,i}}{ln(\frac{\pi L_p}{4 r_p})}\Big]
\end{equation}
Diffusion coefficients $D_{i}$ and $D_{e}$ have been taken from the literature for air \cite{fertig_transport_2001,murphy_transport_1995}, argon \cite{devoto_transport_1967,murphy_transport_1994-1,rat_transport_2002} and helium\cite{murphy_transport_1997}. 
While the voltage current relation  $\frac{d I}{d \varphi}$ and  electron and ion saturation current  $(I_{e,i})_{sat}$ can be measured, the equations \ref{eq:ep:1} and \ref{eq:ep:2} can not be solved analytically. Therefore, the set of equations was iterated for electron temperature $T_e$, ion temperature $T_i$ and electron density $n_e$ until agreement with the measurement values was achieved. 
Furthermore, the influence of the RF-signal from the inductively-heated plasma generator has to be evaluated. If the amplitude of the RF signal is relatively low in comparison to the probe voltage, the bias of the probe signal is low enough to not influence the measurement significantly\cite{oksuz_analysis_2006-1,boschi_effect_1963}. 
For selected measurements, the electron energy distribution function (EEDF) $f(\epsilon = -e V)$ was evaluated \cite{demidov_electric_2002} from the second derivative of the probe current density $\frac{\partial^2 j_e}{\partial V^2}$ as \begin{equation}\label{eq:EEDF}
f(\epsilon = -e V) = - \frac{m_e^2}{2 \pi e^3} \frac{\partial^2 j_e}{\partial V^2}
\end{equation} with the current density being $j_e = e n_e \sqrt{T_e/(2 \pi m_e)} exp(-e V / T_e)$.  For higher electron energies with $\epsilon \geq 3$ \si{\electronvolt}, measurement was not possible due to the limit of the applied probe potential of $V_p = \pm 50$ \si{\volt}. However, the electron density can still be calculated from the EEDF employing the relation \begin{equation}\label{eq:ne_EEDF}
n_e = \int_{0}^{\infty} f(\epsilon) d\epsilon
\end{equation}. 

\section{Operational Parameters of the Facility}

\begin{table}[b]\centering
	\begin{tabular}{ll}
		\hline 
		\hline 
		Parameter	& Range  \\ 
		\hline 
		Operating Frequency $f$& 13.56 \si{\mega\hertz} \\  
		Vacuum Chamber Pressure $p_{vac}$& 0.2 - 400 \si{\pascal} \\  
		RF Power $P_{RF}$& 0.25 - 15 \si{\kilo\watt} \\ 
		Mach Number $Ma$& 0.2 - 1.4 \\ 
		Electron Temperature $T_e$& 0.75 - 2 \si{\electronvolt} \\ 
		Electron Density $n_e$& $10^{18} - 10^{20}$ \si{\meter^{-3}} \\ 
		Generator Efficiency $\eta$& 0.01 - 0.45 \\ 
		Specific Enthalpy $h_s$& 1 - 35 \si{\mega\joule\per\kilogram} \\ 
		Working Gases & Air, Argon, Helium, Nitrogen, Oxygen \\ 
		\hline 
	\end{tabular} 
\caption{Operational Parameters of the IPG6-B facility}\label{table_param}
\end{table}

For the characterization and planning of future experiments within the facility, the previously described diagnostics have been used to determine the operational parameters of the facility. Selected parameters are presented here to give an overview of facility properties and the potential range of applications. For clarity, the important parameters of the facility are listed within Table \ref{table_param}.

\paragraph{Background Vacuum Chamber Pressure}
Especially for the research of electric propulsion but also for the simulation of astrophysical environments, the knowledge of the vacuum chamber pressure is essential. In addition to vacuum chamber pressures, the injector pressures within the discharge tube have been measured to gain information on future achievable nozzle exit velocities. The vacuum pump is capable of reaching a base vacuum chamber pressure of 0.2 Pa. In Fig. 4, the back vacuum chamber pressure is displayed as a function of the volume flow for air, argon, helium and nitrogen. It can be seen that the vacuum chamber pressure is a linear function of the volume flow rate independent of the gas type, while the injector pressure has a dependence on plasma power and gas type. This is most likely due to the position of the vacuum chamber pressure gauge, where most of the plasma has already recombined and cooled down. Therefore the effect of dissociation and temperature which is seen for the injector pressure is not observed for the vacuum chamber pressure.
\begin{figure}
	\centering
	\includegraphics[width=0.725\linewidth]{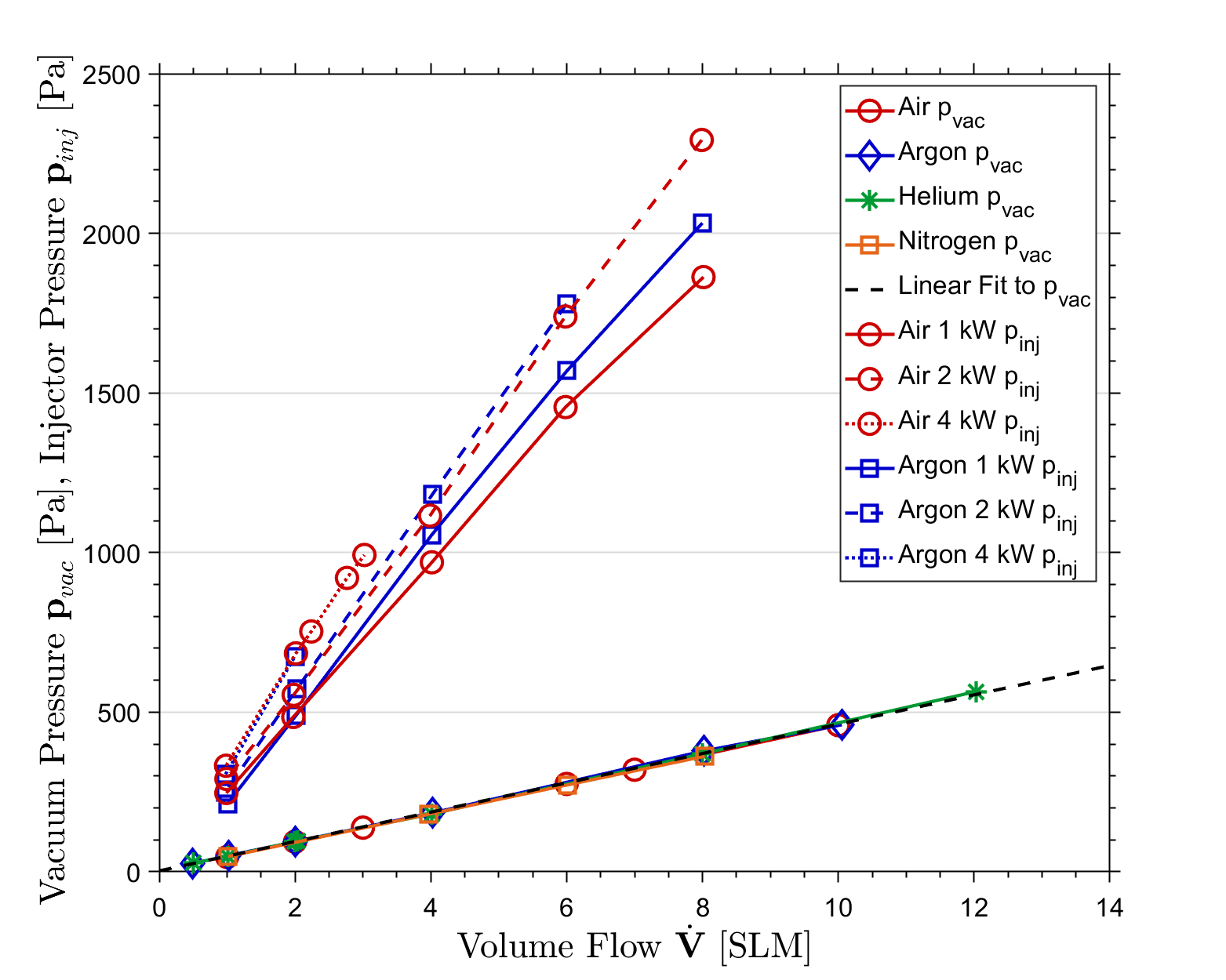}
	\caption{Vacuum chamber pressure $p_{vac}$ and injector pressure $p_{inj}$ of the facility as a function of the Volume Flow Rate $\dot{V}_{gas}$ for air (red), Argon (blue), Helium (green) and Nitrogen (orange) at different coupled RF powers $P_{RF}$. The injector pressure is measured during operation with a 10mm nozzle. The same colours for different gas type apply to all other figures displayed. }
	\label{fig:vacpressure}
\end{figure}

\begin{figure}
	\centering
	\includegraphics[width=0.725\linewidth]{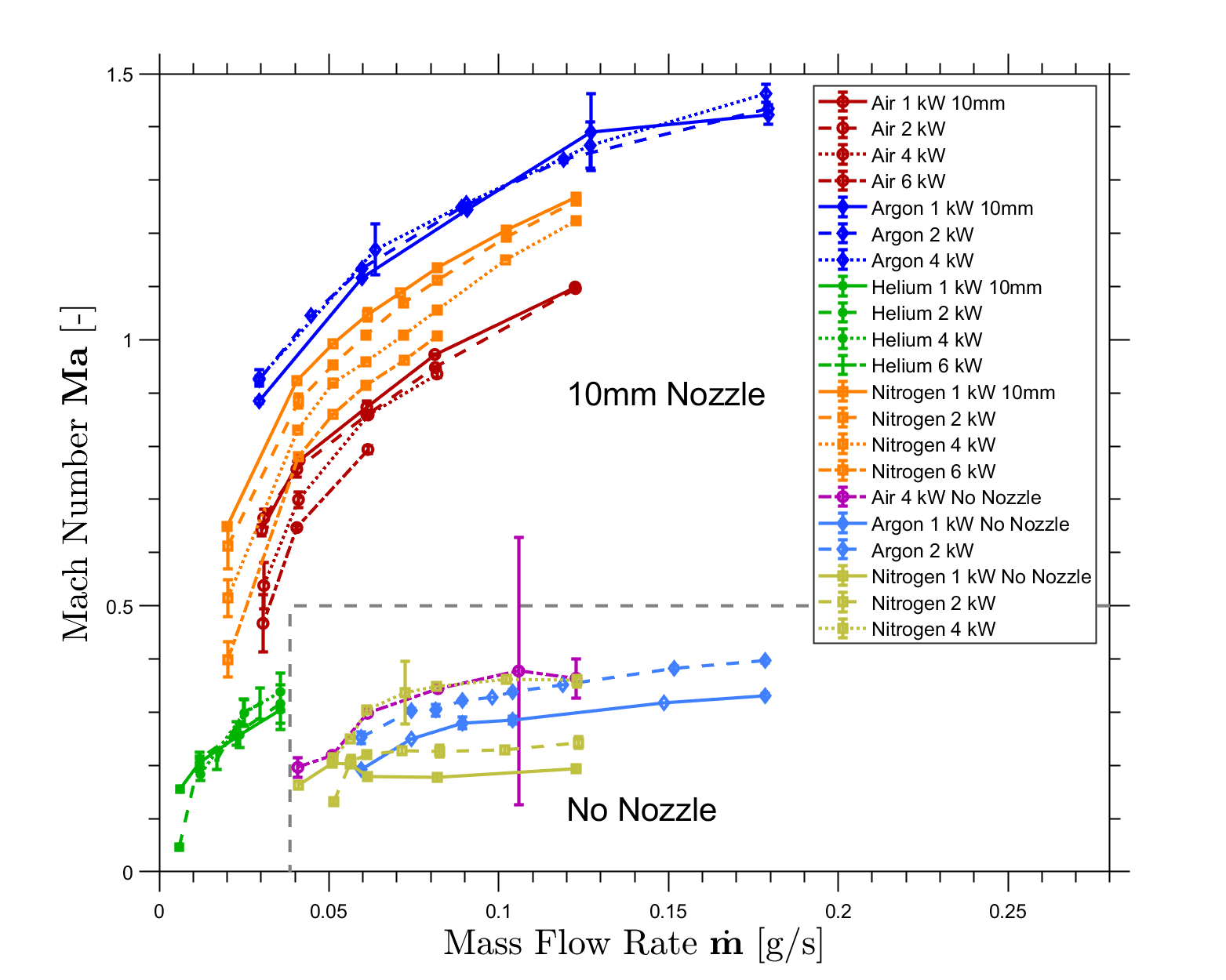}
	\caption{Axial Mach Number $Ma$ as a function of mass flow rate $\dot{m} $ for Air (red and pink), Argon (blue), Helium (green) and Nitrogen (orange and yellow) for different RF Powers  $P_{RF}$ in the configuration with a 10mm nozzle at a distance of 100mm from the nozzle. The vacuum chamber pressure is a function of volume flow as shown in Fig. \ref{fig:vacpressure}. }
	\label{fig:machnumbers}
\end{figure}

\paragraph{Mach numbers}
To compare flow conditions within the facility with conditions appearing in applications such as atmospheric entry flows or astrophysical plasmas as well as to determine the absolute flow velocity, the Mach number of the flow was determined using a pitot probe. It has been found that the facility can operate across a wide range of Mach numbers, creating sub- and supersonic flows. In Fig. 5, the flow Mach number at a distance of z = 100 mm from the (divergent) nozzle exit of a nozzle with a throat diameter of 10 mm is shown as a function of the volume flow rate into the injector of the plasma generator. As shown, the Mach number has a strong dependence on the volume flow rate and varies significantly for different gas types. The highest Mach numbers are achieved using Argon while the lowest are achieved using Helium. It can also be seen in Fig. \ref{fig:machnumbers}, that the measured Mach number $Ma$ decreases for an increase RF Power $P_{RF}$, most clearly observed in the results for Air and Nitrogen. This is most likely explained by a combination of different effects: The Mach number is for constant velocity inversely proportional to the temperature $Ma \propto \sqrt{T}^{-1}$, therefore the Mach Number for the same gas mass flow rate $\dot{m}_{gas}$ will be lower at higher RF Powers $P_{RF}$ due to the increase in temperature. Additionally, the gas is not expanded ideally in the divergent nozzle (as visible by the Mach disks in Fig. \ref{fig:facilityoperation}), and therefore not all of the thermal energy of the plasma is converted into kinetic energy. This effect might be lowered by the extension of the nozzle with a divergent section.

\paragraph{Plasma power, Efficiency \& Enthalpy}
For application of the facility as a plasma wind tunnel for re-entry flows or the study of high-temperature materials such as those used as wall materials in fusion devices, knowledge of plasma flow enthalpies is essential. To compare the flow conditions within the facility with the conditions appearing in other applications, a cavity calorimeter has been used to measure the plasma power. In Fig. 6, the overall plasma generator efficiency $\eta$, defined as as the ratio $\eta=P_{Cal}/P_{RF}$ between measured calorimeter power $P_{Cal}$ and coupled RF Power $P_{RF}$ is shown for different gases as a function of the RF Power. It can be seen, that for most gases, generator efficiency drops significantly with an increase of the RF power, while it rises for Helium. The efficiency of the generator is significantly higher for Air and Nitrogen, reaching 45\% while dropping below 10\% for the noble gases. In Fig. 7, the specific enthalpy, evaluated from calorimeter measurements and defined as $h_s = P_{Cal}/\dot{m}_{gas}$, is displayed as a function of the RF Power. As shown, the specific enthalpy rises significantly, despite the drop in generator efficiency for higher powers. As such, flow enthalpies of 32 MJ/kg can be reached within the IPG6-B facility, similar to the flow enthalpy encountered during re-entry from a low-earth-orbit.

\begin{figure}
	\centering
	\includegraphics[width=0.725\linewidth]{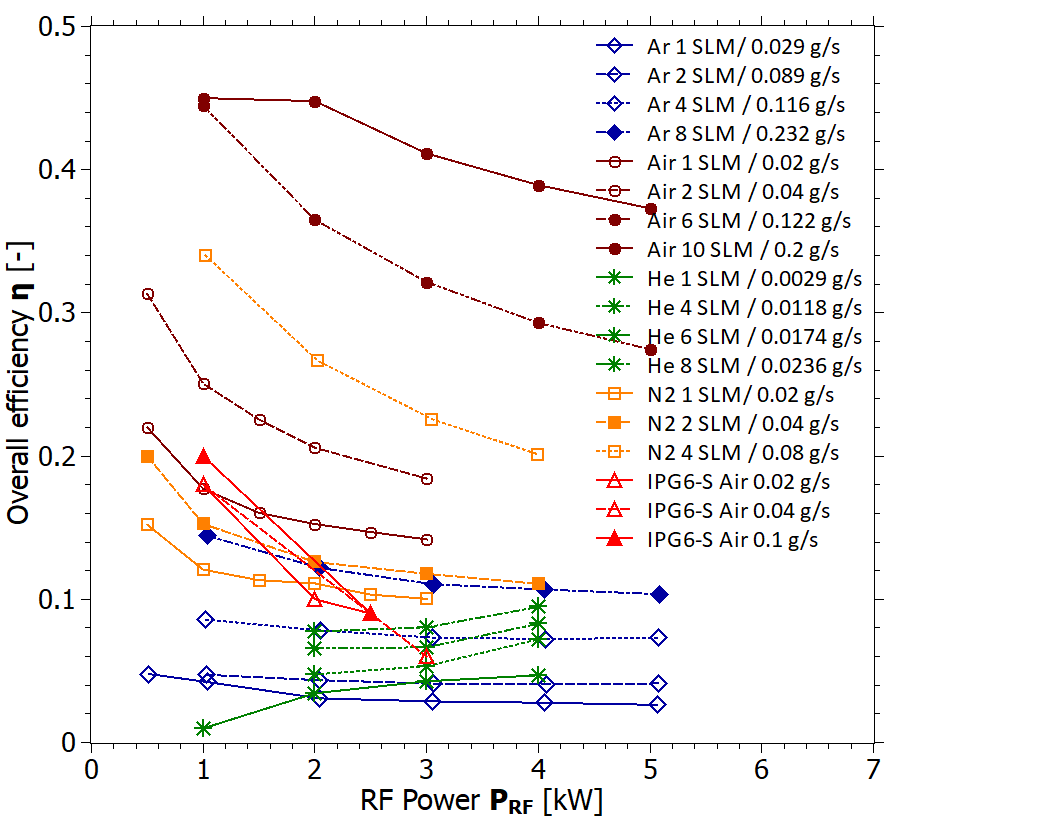}
	\caption{Overall efficiency $\eta$ of the plasma generator as a function of the coupled RF power for air, Argon, Helium and Nitrogen at different volume flow rates in operation without a nozzle calculated from calorimeter measurements of the plasma power at a distance of 25 mm from the nozzle exit, December 2017 (Air, Argon, Nitrogen) and October 2018 (Helium). The bright red line  displays measurements from the IPG6-S facility for air. The vacuum chamber pressure is a function of volume flow as displayed in Fig. \ref{fig:vacpressure}.  }
	\label{fig:efficiency}
\end{figure}

\begin{figure}
	\centering
	\includegraphics[width=0.725\linewidth]{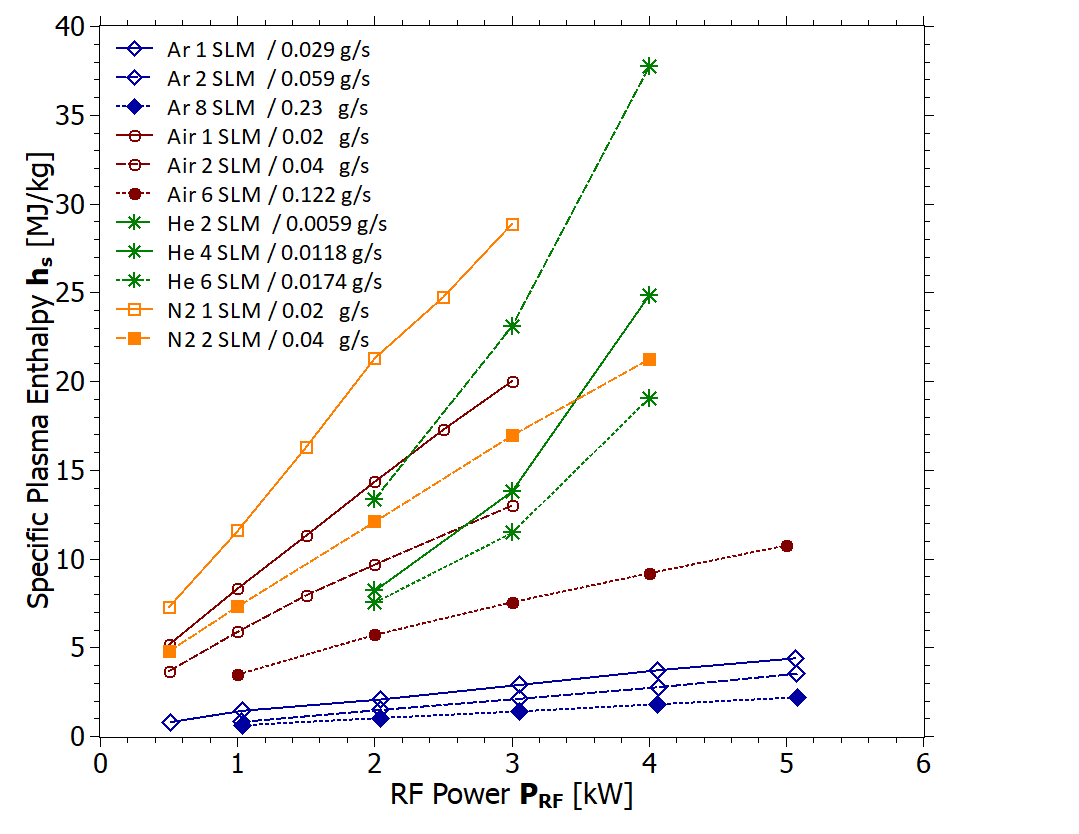}
	\caption{Specific plasma enthalpy $h_s$ as a function of RF input power $P_{RF}$ for air, Argon, Helium and Nitrogen calculated from the same measurement campaign as Fig. \ref{fig:efficiency}.}
	\label{fig:enthalpy}
\end{figure}

\paragraph{Electron Temperature}
\begin{figure}[h]
	\centering
	\includegraphics[width=0.75\linewidth]{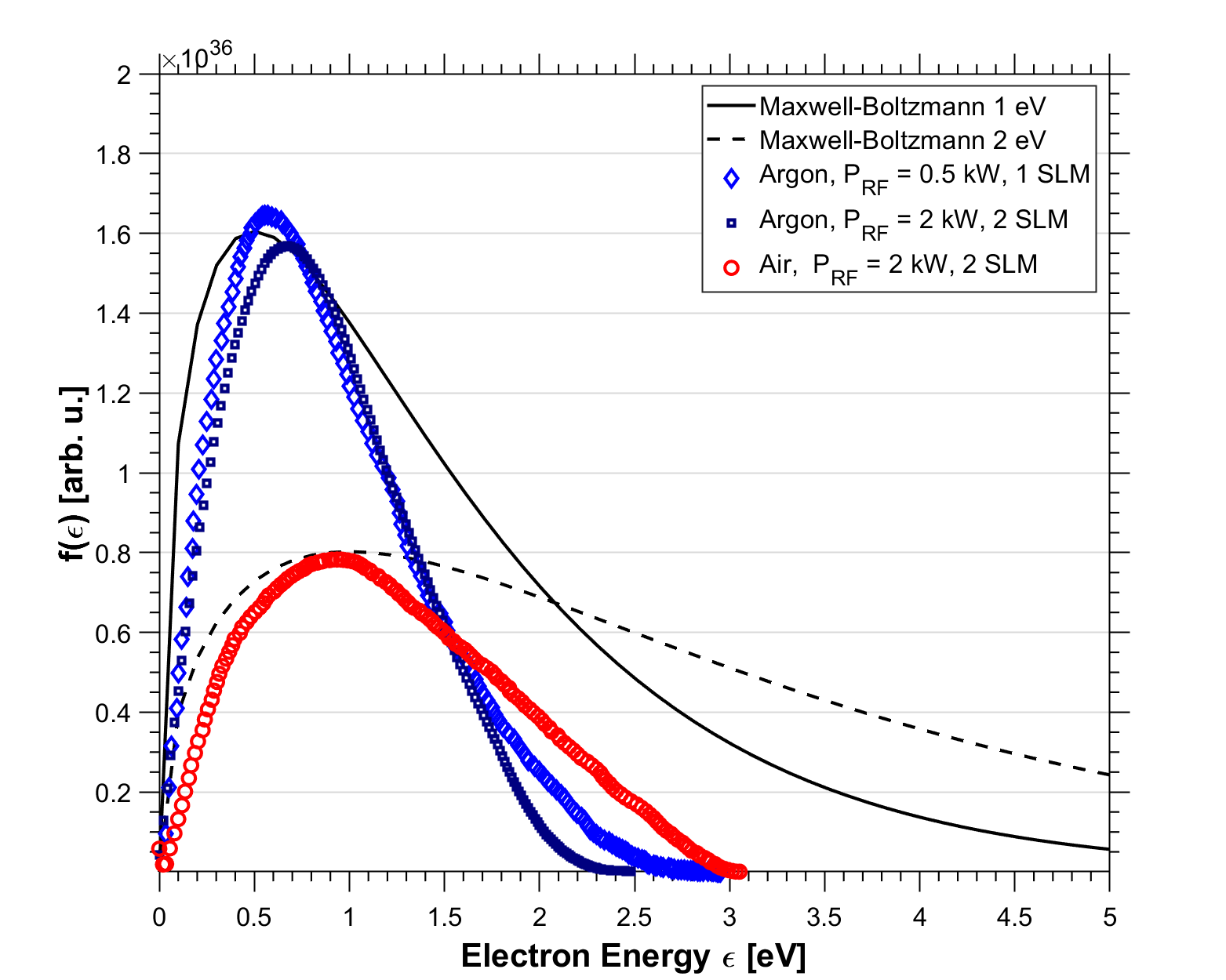}
	\caption{The electron Energy distribution function (EEDF) for air (red) and Argon (blue) at a RF power of 0.5 and 2 kW in operation with a 40mm nozzle  in comparison to a Maxwell-Boltzmann distribution for $T_e = 1$ \si{\electronvolt}  (solid black) and  $T_e = 2$ \si{\electronvolt}  (dashed black) at a pressure of 50 Pa at a volume flow rate of 1 SLM and a pressure of 100 Pa at a volume flow rate of 2 SLM measured in a distance of 300mm from the nozzle exit, October 2018. The measured electron density was $n_e \approx 6 \cdot 10^{19}$\si{\meter^{-3}}}
	\label{fig:EEDF}
\end{figure}

To determine plasma properties such as electron temperature, electron density and ion temperature, electrostatic single probe measurements have been conducted within the facility \cite{schmidt_electrostatic_2018}. Although the relatively high pressures and low electron temperatures within the IPG6-B made it necessary to apply continuum electrostatic probe theory, measurement of the EEDF was still possible for certain working points. In Fig. \ref{fig:EEDF}, the measured EEDF, calculated using equation \ref{eq:EEDF}, is shown for air and argon in comparison to a Maxwell-Boltzmann distribution for electron temperatures of $T_e = 1$ \si{\electronvolt} and  $T_e = 2$ \si{\electronvolt}. 
For both measurements, it can be seen that there is a peak of the EEDF at low electron energies around $\varepsilon = 1$ \si{\electronvolt}, while there are no results for electron energies above $\varepsilon \geq 3$ \si{\electronvolt} and the EEDF decreases strongly between energies of $1$ \si{\electronvolt} and $3$ \si{\electronvolt}. Since the resulting EEDFs do not look Maxwellian, an explanation is necessary:  Due to limitations of the signal generator and amplifier, it is not possible to apply a higher probe potential for the measurement of the EEDF. Therefore, information on the high-energy tail of the EEDF remains unknown, since no potential can be applied to gain information about the electrons with an energy above $\varepsilon \geq 3$ \si{\electronvolt}. 
Still, the results within the low-energy region of the EEDF are in agreement with other measurements of the EEDF in inductively-coupled plasmas. For argon, it has been shown experimentally \cite{godyak_abnormally_1990,godyak_measurement_1992,godyak_electron_2002,boffard_optical_2010} and numerically \cite{yonemura_electron_2001,meige_electron_2006,hagelaar_coulomb_2015} that for pressures above $p_{vac} \geq 10$ \si{\pascal} the high-energy tail of the EEDF is depleted, while the low-energy region increases due to change of the plasma heating mode from stochastic to collisional heating. The low-energy regions of the EEDF in these works agree well with the measured EEDF shown in Fig. \ref{fig:EEDF}. For diatomic gases, specifically nitrogen \cite{singh_measurements_2000,schwabedissen_langmuir_1997}, a gap within EEDF for electron energies of $\varepsilon = 3$ \si{\electronvolt} has been observed due to the presence of
resonant electron–molecule vibrational excitation cross sections, which is in agreement with the behaviour of the EEDF observed in Fig. \ref{fig:EEDF}.
For future measurements of the EEDF in the IPG6-B, the use of a different signal amplifier would bring a great advancement to gain further knowledge about the EEDF at high electron energies. Nonetheless, from these measurements average electron temperatures of $T_e = 0.75 - 2 $ \si{\electronvolt} and electron densities of $n_e = 10^{18} - 10^{20} $ \si{\meter^{-3}} can be estimated depending on gas type, flow rate and pressure as well as by calculating it from the EEDF employing equation \ref{eq:ne_EEDF} within the previously discussed limits.

\section{Charged Dust as a Diagnostic}
\begin{figure}[t]
	\centering
	\includegraphics[width=\linewidth]{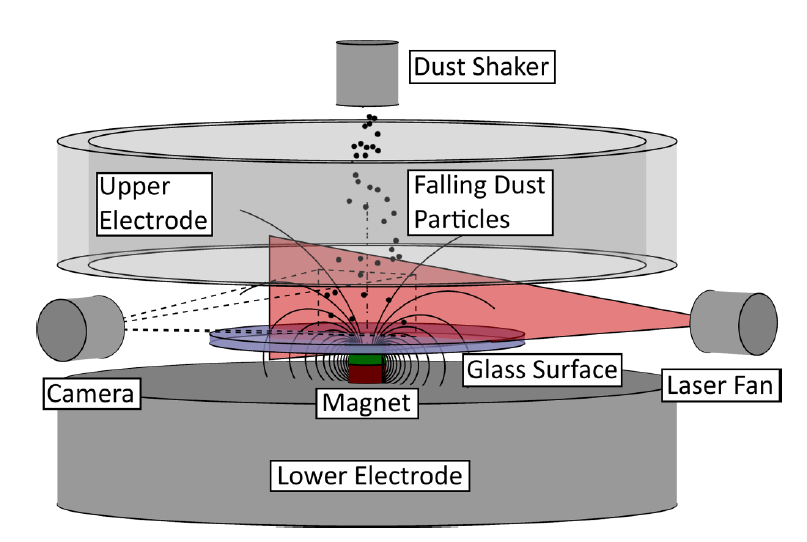}
	\caption{Schematic of  magnetic field mapping in a quiescent plasma in a GEC Reference Cell at CASPER \cite{dropmann_analysis_2015-1}. 11.4\si{\micro\meter} melamine-formaldehyde dust particles are dropped on an electrode and interact with the magnetic field applied by a permanent magnet.}
	\label{fig:dustshaker}
\end{figure}

\begin{figure}
	\centering
	\includegraphics[width=\linewidth]{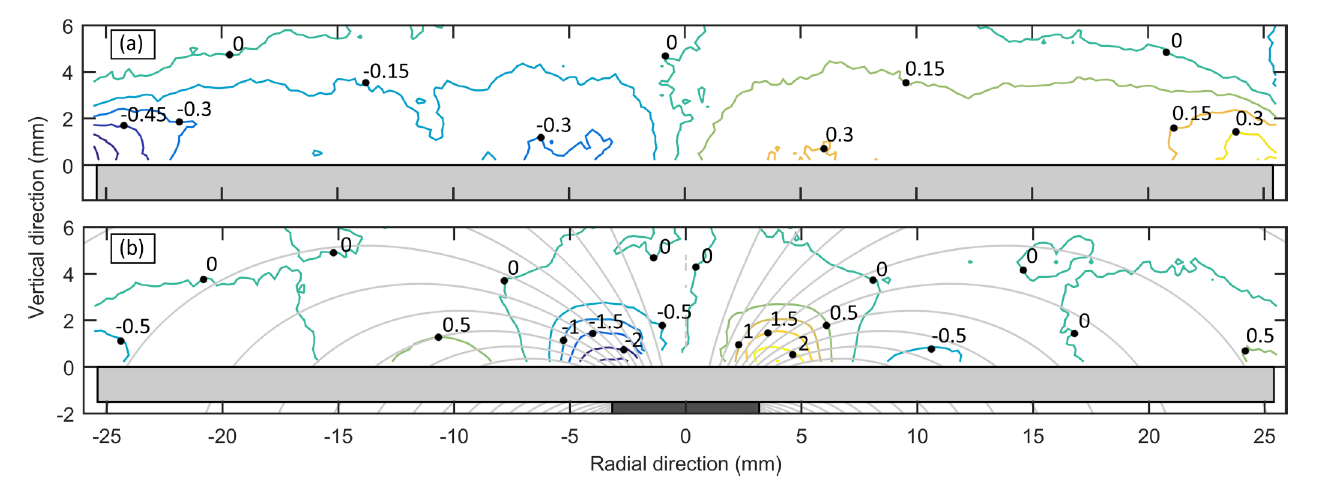}
	\caption{Radial Acceleration of particles in a quiescent plasma in a GEC Referemce Cell at CASPER \cite{dropmann_analysis_2015-1} with (lower image) and without (upper image) applied magnetic field. The magnetic field lines are plotted in grey. }
	\label{fig:dust_MD}
\end{figure}

In a multiphase flow, a particle interacts with the gas through drag forces. Within a complex plasma, since the dust particles are charged due to free electrons within the dusty plasma system, additional interaction forces rapidly become important. Among others, these include the interaction of the (negatively charged) particles with the plasma, the creation of ion wakes, the interaction of the particles with one another and the interaction of the particles with applied or self-induced electric and magnetic fields. These interactions can be described by
\begin{equation}\label{eq:dust}
m_p \frac{\partial \vec{v}}{\partial t} = F_{E} + F_{B} + F_{D,n} + F_{D,i} \mathrm{.}
\end{equation}
In equation \ref{eq:dust}, $F_{E} = q \cdot \vec{E}$ describes the electrostatic and $F_{B} = q \cdot \vec{v} \times \vec{B}$ the electromagnetic forces acting upon a dust particle with mass $m_p$, velocity $\vec{v}$ and charge $q$ in an electric field $\vec{E}$ or magnetic field $\vec{B}$, while $F_{D,n}$ and $F_{D,i}$ describe the drag by neutrals and the drag by ions, respectively. Since charged dust particles are directly affected by the electric field, the charged particles (as well as the electrons and ions depending on the magnitude of the field) can be ‘focused’ along the magnetic field lines if a magnetic field is applied. Due to their mass, micrometer sized particles must be exposed to a magnetic field of at least 1 \si{\tesla} in order to observe gyrations. Thus, an applied electric field with magnitude of 10-20 \si{\kilo\volt\per\meter} can significantly influence them. Even if such fields cannot be achieved, lower field strengths can still affect the plasma and therefore the entrained dust due to its entrenched nature and interaction with the plasma. As such, this effect can be used to experimentally study the magneto hydrodynamic interaction of the plasma with applied and self-induced fields providing an interesting opportunity to ‘map’ the plasma at much lower field strengths. While the dust-field interaction has been used to study the electric field within the ion-wake of a particle \cite{chen_ion-wake_2016}, map the plasma potential \cite{scott_mapping_2019,wolter_micro-particles_2009} and study the RF sheath \cite{samarian_dust_2005,samarian_dust_2009} within a glass box, the interaction of dust as a flow diagnostic for the interaction of plasma flows with magnetic fields has not been studied very extensively in experiments yet.
\paragraph{Prior experimental work in quiescent plasmas}
This method has already been applied to stationary plasmas interacting with the field created by permanent magnets\cite{dropmann_analysis_2015-1}. In the experiment shown in Fig \ref{fig:dustshaker}, 11.4\si{\micro\meter} melamine-formaldehyde (MF) dust particles with a density of 1.51 \si{\gram\per\centi\meter^3}are dropped onto a planar, circular electrode creating a capacitively-coupled RF discharge at a plasma power of 11 \si{\watt} and a pressure of 5.3 \si{\pascal}. To examine the influence of the magnetic fields on the particle trajectories, a  cylindrical NdFeB magnet of grade 40 SH with a magnetic remanence of 1.24-1.28 \si{\tesla}, 6.35 mm in both length and diameter, was placed on the lower electrode beneath a glass plate of 1.5mm thickness and 50.8 mm diameter. Particles were illuminated using a laser fan and a Photron 1024 high speed camera was used to track the trajectories of single particles. Evaluating the forces acting on the particles as described by equation \ref{eq:dust}, the radial accelerations of the particles were evaluated. As seen, in Fig. \ref{fig:dust_MD}, the difference in radial acceleration between experiments, where a magnetic field is present, is significant. As such, this can be used to map the magnetic field and hopefully extended to allow more complex geometries.

\paragraph{Dust within plasma flows}

\begin{figure}
	\centering
	\includegraphics[width=\linewidth]{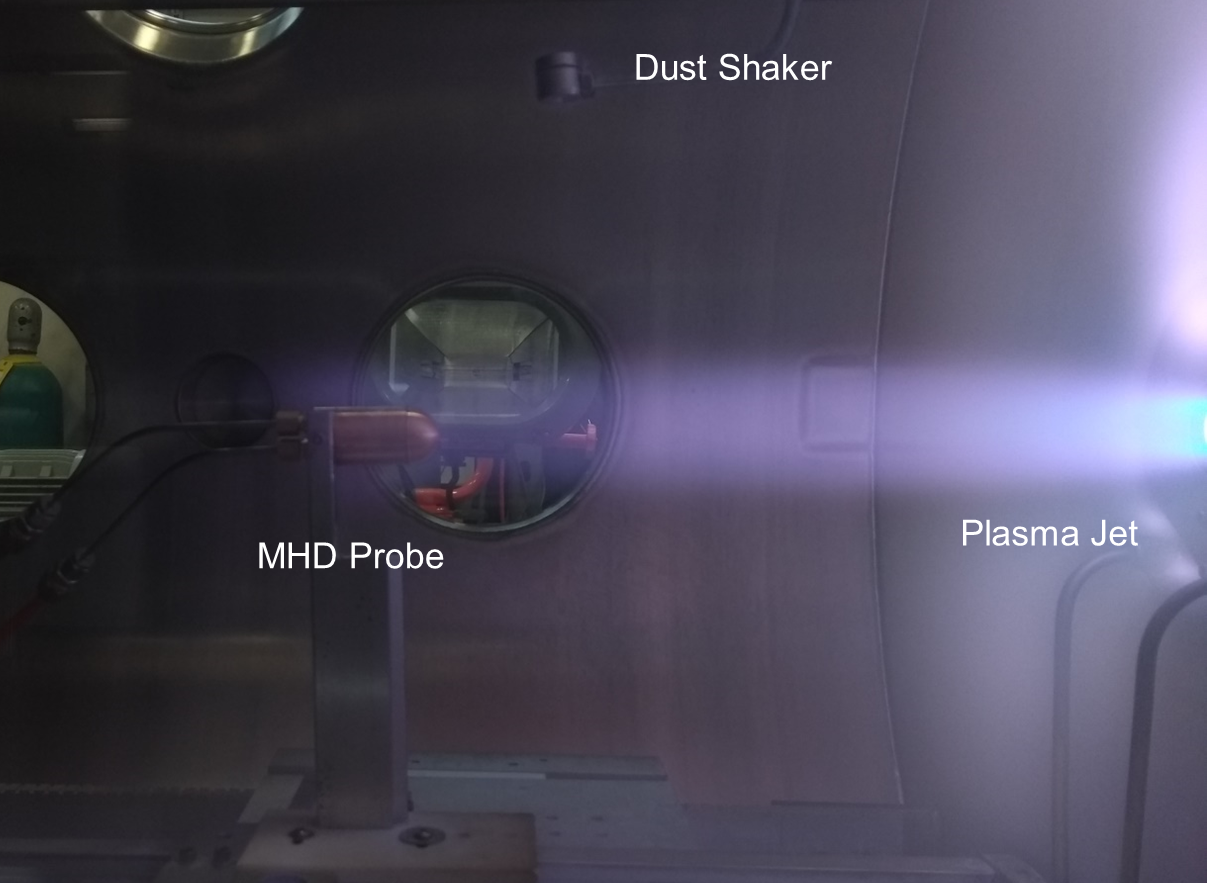}
	\caption{Picture of an experiment in which dust particles are injected into a plasma flow before interacting with a spherical obstacle at a Mach number of Ma = 0.3, a volume flow of $V = 2.5$ SLM, pressure of $p_{vac} = 90$ Pa and a distance of 300 mm from the nozzle. The coupled plasma power was $P_{RF} = 1000$ \si{\watt}. }
	\label{fig:dustMHD}
\end{figure}

\begin{figure}
	\centering
	\includegraphics[width=\linewidth]{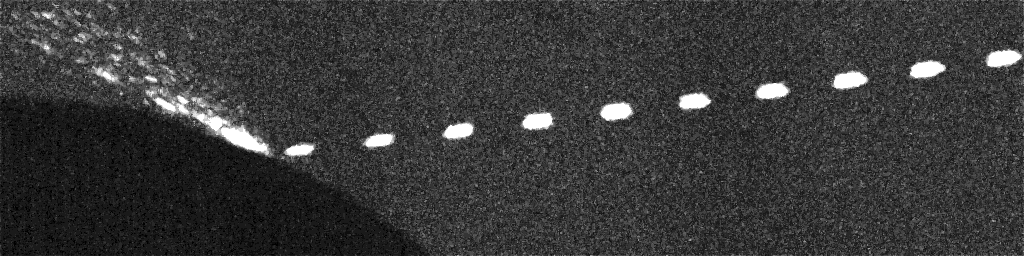}
	\caption{Time lapse image of 20 \si{\micro\meter} alumina dust entrained in a 0.5 kW Argon plasma jet at a Mach number of Ma = 0.3, with a volume flow of $V = 2.5$ SLM, pressure of $p_{vac} = 90$ Pa and a distance of 300 mm from the nozzle interacting with a spherical obstacle with a diameter of $d = 25$ \si{\milli\meter}. The agglomerated particle is shown colliding with the obstacle and the trajectories of single particles after the breakup are visible. The flow is coming from the right. Exposure time: 10\si{\micro\second}, Framerate: 3000 fps. A picture of this experiment is shown in \ref{fig:dustMHD}. March 2020 }
	\label{fig:dust2}
\end{figure}

Another representative application in low-temperature plasma is the study of the interaction between particles and fields in magneto hydrodynamic (MHD) flows. In a recent experiment, a water-cooled blunt body 'MHD probe' with a radius of 12.5 \si{\milli\meter} was placed within an argon plasma jet created within the IPG6-B facility as shown in Fig \ref{fig:dustMHD}. Within the inside of the water-cooled probe, one or two NdFeB permanent magnets can be placed to create a surface magnetic flux density of up to $B = 124$ \si{\milli\tesla} at the stagnation point of the probe. In this experiment, 20 \si{\micro\meter} alumina ($Al_{2}O_{3}$) dust was dropped using a manually operated dust shaker placed above the plasma jet. The dust was illuminated using background lighting and the dust trajectories were recorded using a high speed Photron FastCam 100K camera. A time lapse image of the trajectories of an agglomerated particle colliding with the blunt body before separating into smaller single particles is shown in Fig. \ref{fig:dust2} demonstrating the capacity of the imaging system and the possibility of observing even micrometer sized particles within the flow. The trajectories were further evaluated using particle tracking and particle image velocimetry (PIV) methods. Employing this technique, dust can be used to measure not only the flow velocity fields within a plasma jet, but also to map the electric and magnetic fields. Taken together, this is therefore a promising new diagnostic technique to be established within future research. Compared to conventional PIV used as a diagnostic for gas flows, with charged dust, crucial information on electric and magnetic fields can also be achieved using the described method. This should allow the study of more complex obstacle and magnetic field geometries, were the field geometry might now be known initially.

\section{Potential Future Applications of the facility}
\begin{figure}[t]
	\centering
	\includegraphics[width=1\linewidth]{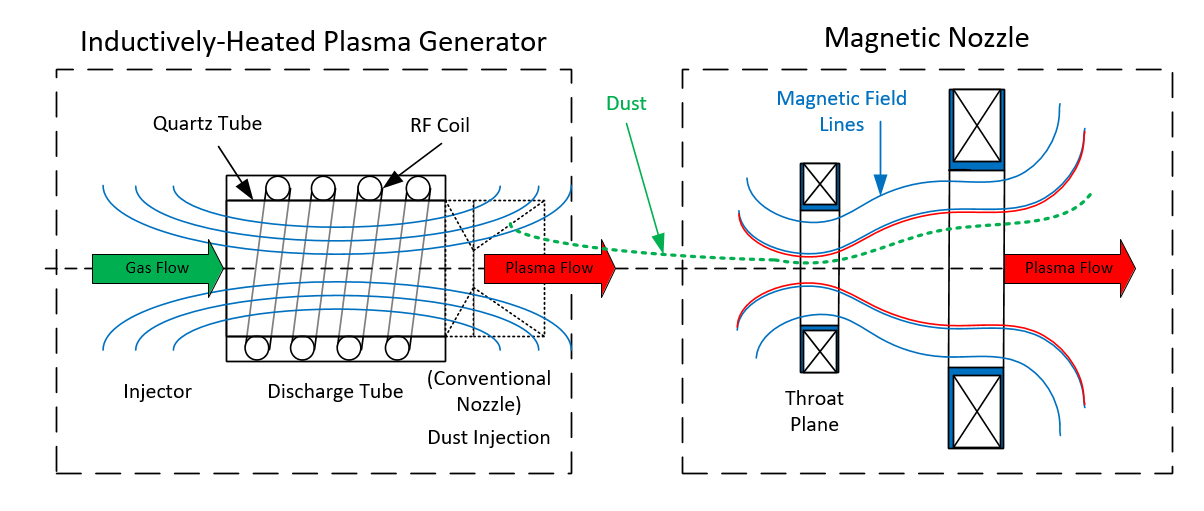}
	\caption{Concept for a future magnetic nozzle experiment in which dust is entrained within a plasma jet created by the IPG6-B facility. The plasma jet containing the dust is directed into the magnetic field geometry created using an electromagnet and the trajectories of the dust are studied using optical methods. The flux density of the magnetic nozzle can reach between  0.1 - 1 \si{\tesla}.}
	\label{fig:experiment}
\end{figure}

Due to the flexibility of the IPG6-B facility, a wide range of applications is considered feasible across various fields of physics and engineering. These can be roughly separated into three categories:
\paragraph{Engineering research}
This includes the study of plasmas used in electric propulsion or occurring during atmospheric entry of spacecraft as well as the study of wall materials for magnetic confinement fusion devices such as Tokamaks and Stellarators. It can be seen, that  the specific flow enthalpy  reaches $35$ \si{\mega\joule\per\kilogram} for helium and for air $20$ \si{\mega\joule\per\kilogram} respectively, which becomes comparable to the parameters observed during re-entry from a low-earth-orbit for a velocity of $ v_{\inf} = 8 $ \si{\kilo\meter\per\second}, where the specific enthalpy of the flow is approximately $h_{s,\inf} \approx v_{\inf}^2 / 2 = 32$ \si{\mega\joule\per\kilogram}\cite{kolesnikov_similarity_2016}. This partially demonstrates the ability to reproduce sub- and supersonic high enthalpy flows as they occur during re-entry as well as similar high-temperature conditions related to hypersonic flight and combustion. The interaction of a plasma field with magnetic fields applied to the flow is of special interest for problems concerning magnetic heat flux mitigation.
\paragraph{Experimental studies of magnetic nozzles}
Within a magnetic nozzle, the plasma flow becomes confined by the surrounding magnetic field. It has been shown that these plasma flows follow the converging and diverging magnetic field lines in a manner similar to the compression and expansion observed in a de Laval-Nozzle \cite{andersen_continuous_1969,ahedo_two-dimensional_2012}. In this case, the magnetic compression is created primarily due to the electrons being magnetized by the field and then forced to follow the magnetic field lines \cite{breizman_magnetic_2008}. The ions are only partially magnetized and follow the electric field created by charge separation of the electrons due to the condition of quasi-neutrality\cite{arefiev_magnetohydrodynamic_2005}. At a specific detachment point within the magnetic nozzle, the magnetic forces become too weak to confine the ions, which are then accelerated away from the nozzle. The electrons now follow the ions due to ambipolar diffusion driven by the electric field created by charge separation. This process is called plasma detachment \cite{breizman_magnetic_2008} and is the primary mechanism for thrust creation in a magnetic nozzle. As conventional Laval-nozzles used for the acceleration of gas and plasma flows are limited in the achievable exit Mach-number, the use of a magnetic nozzle seems promising for future plasma science, plasma processing and space propulsion, while also verifying an entirely new diagnostic method for future fusion devices. A possible experiment to study a magnetic nozzle is shown in Fig \ref{fig:experiment}. Within the IPG6-B facility, a plasma is created and dust is injected into the plasma jet to be used as a diagnostic. The dust entrenched in the plasma flow then enters the magnetic field geometry of the magnetic nozzle, created either by an electromagnetic field using a conventional or super-conducting coil or permanent magnets. Depending on the source used for the magnetic field, the magnetic flux density can range between 0.1 - 1 \si{\tesla}. The interaction of the dust and plasma within the magnetic nozzle geometry can then be studied using the above mentioned optical diagnostics. This can give information on field geometries and forces within the magnetic nozzle, which becomes crucial as magnetic nozzles have become a feasible method of thrust creation for advanced electric propulsion devices such as helicon-wave based thrusters \cite{romano_inductive_2019-1,romano_inductive_2019-2,takahashi_helicon-type_2019}, and a more detailed understanding of the processes within magnetic nozzles can lead to further understanding and improvement of future magnetic nozzles as well as other technical applications \cite{andersen_continuous_1969}, in which plasma-magnetic field interaction can be found such as Hall-effect thrusters or in magnetic confinement fusion devices. 
As these engineering problems consist of more complex or unknown field geometries, the extension of the already presented theory on the dust magnetic field interaction becomes necessary to use charged dust to study these effects.

\paragraph{Astrophysics \& Space Physics}
Due to the wide range of pressures and plasma conditions achievable within the facility, it offers the possibility of studying several astrophysical phenomena, such as the interaction of the plasma with magnetic fields (i.e. those observed in Earth’s magnetosphere) or the accretion of fine grained dust rims on chondrules within the protoplanetary disk that formed the planets within our solar system \cite{schmidt_concept_2019,schmidt_behavior_2020}.
\paragraph{Fundamental Physics}
Further applications of the facility are directed toward the study of fundamental physical phenomena. A special focus of the IPG6-B at CASPER is dusty plasma. Charged dust can not only be used as a diagnostic method for plasma, but can also interact with the plasma in various ways that are of interest for further research,  one example is the formation of dusty plasma chains and crystals due to electric potential interaction \cite{fortov_dusty_2004}. Furthermore, plasmas created within the IPG6-B show some similarity to plasma observed close to the wall and the divertor region of magnetic confinement fusion devices such as Tokamaks \cite{bastykova_simulation_2019}. As the production and interaction of dust at the wall of a fusion plasma is of high interest, this facility gives therefore the possibility to study such plasma conditions in a more easily accessible environment.
\section{Conclusions and future research}
Details of characterization of the IPG6-B over the past years \cite{zhukhovitskii_electrical_2015,dropmann_new_2013-1,koch_setup_2012} as well as several possible applications have been proposed within this article. These cannot only increase research in fundamental field of physics, but may also support and extend research regarding the development of future electric propulsion concepts for space applications, especially contact-less RF thrusters. Additionally, the facility is not only able to operate at a relatively low cost, it also has the unique ability to use charged dust as a diagnostic method for phenomena that occur within electric propulsion devices. A further study of those phenomena could deepen the understanding of effects occurring within magnetic nozzles, magnetic confinement and Hall-effect thrusters as well as other applications in which plasma-magnetic field interactions play a role and therefore lead to optimization and improvement of future technology enabling new missions and mission concepts. While to date only problems of relative simple geometry have been studied, this research will be extended onto more complex problems as they occur in the recent engineering challenges discussed above. Future research within the facility will not only include the study of fundamental phenomena of plasmas and further characterization, but also the improvement of diagnostics which can be used universally in plasma physics.

\section*{References}

\bibliography{BibActAst}

\end{document}